\documentclass[sn-mathphys]{sn-jnl}% Math and Physical Sciences Reference Style
\jyear{2021}%

\theoremstyle{thmstyleone}%
\theoremstyle{thmstyletwo}%

\theoremstyle{thmstylethree}%

\begin{document}

\title[Modeling of neurotransmitter release]{Quantifying constraints determining independent activation on NMDA receptors mediated currents from evoked and spontaneous synaptic transmission at an individual synapse}


\author*[1,2]{\fnm{Sat byul} \sur{Seo}}\email{sbseo@kyungnam.ac.kr}

\author[2]{\fnm{Jianzhong} \sur{Su}}\email{su@uta.edu}


\affil*[1]{\orgdiv{Department of Mathematics Education}, \orgname{Kyungnam University}, \orgaddress{ \city{Changwon-si}, \postcode{51767}, \state{Gyeongsangnam-do}, \country{Republic of Korea}}}

\affil[2]{\orgdiv{Department of Mathematics}, \orgname{University of Texas at Arlington}, \orgaddress{\city{Arlington}, \postcode{76019}, \state{Texas}, \country{United States}}}

%\affil[3]{\orgdiv{Department}, \orgname{Organization}, \orgaddress{\street{Street}, \city{City}, \postcode{610101}, \state{State}, \country{Country}}}


\abstract{A synapse acts on neural transmission through a chemical process called synapses fusion between pre-synaptic and post-synaptic terminals. Presynaptic terminals release neurotransmitters either in response to action potential or spontaneously independent of presynaptic activity.  However, it is still unclear the mechanism of evoked and spontaneous neurotransmission that activate on postsynaptic terminals. To address this question, we examined the possibility that spontaneous and evoked neurotransmissions using mathematical simulations. We aimed to address the biophysical constraints that may determine independent activation on N-methyl-D-asparate (NMDA) receptor mediated currents in response to evoked and spontaneous glutamate molecules releases. In order to identify the spatial relation between spontaneous and evoked glutamate release, we considered quantitative factors, such as size of synapses, inhomogeneity of diffusion mobility, geometry of synaptic cleft, and release rate of neurotransmitter. Simulation results showed that as a synaptic size is smaller and if the cleft space is more cohesive in the peripheral area than the centre area, then there is high possibility of having crosstalk of two signals released from center and edge. When a synaptic size is larger, the cleft space is more affinity in the central area than the external area, and if the geometry of fusion has a narrower space, then those produce more chances of independence of two modes of currents released from center and edge. The computed results match well with existing experimental findings and serve as a road map for further exploration to identify independence of evoked and spontaneous releases.}

\keywords{synaptic transmission, neurotransmission, NMDA receptor, mathematical modeling, differential equations}



\maketitle

\section{Introduction}\label{sec1}

 A synapse  consists of the three components, a presynaptic neuron or terminal,  a postsynaptic neuron or terminal, and  a synaptic cleft. Networks of neurons and synapse play a key role of communication of electric signals of brain, and are responsible for most of brain functionality. Chemical synapses between neurons, are the main channels of information flow and storage in the brain. Synaptic transmission between neurons is involved in most of what the brain does. In mouse cortex are, synaptic neuropil is 84\% in terms of volume\cite{Schuz1989}. When a neuron is active, an electrical impulse travels down its nerve fiber and causes the release of chemical neurotransmitters from its terminal on the presynaptic neuron. Presynaptic terminals contain pools of synaptic vessels. They are small membrane bounded organelles. These vesicles are filled with neurotransmitters mainly in the form of glutamate molecules. When neurotransmitters, for example, glutamate molecules are released from the presynaptic neuron, they diffuse into the cleft, and then the neurotransmitters spread out to a narrow space between pre and post synaptic neurons. The gap between pre and post synaptic terminals is about 20nm wide and is called a synaptic cleft. The synaptic cleft consists of fluids, proteins and other molecular obstacles. The some of released neurotransmitters in the cleft bind to receptors on the postsynaptic neuron\cite{ledoux2003synaptic}. These chemical neurotransmitters then produce secondary currents in the postsynaptic neuron. 

The random synaptic release events typically activate receptors within a single postsynaptic site and give rise to miniature postsynaptic currents, and therefore they have been extremely instrumental in analysis of unitary properties of neurotransmission. In 1994, Murphy and colleagues found that spontaneous miniature glutamate release modulates postsynaptic enzyme activity\cite{Kenny1994}. Sutton and colleagues showed that minis keep resting protein synthesis in check and respond to stimuli that strengthen synapses by blocking minis and increasing dendritic protein synthesis\cite{Sutton2004}. Spontaneous neurotransmission has been mentioned a homeostatic form of synaptic plasticity and induction of synaptic scaling. Spontaneous neurotransmission has an independent role in neuronal communication that is distinct from that of evoked release\cite{Kavalali2014} However, the process of spontaneous neurotransmitter release is still unclear. It has been questioned whether spontaneous release events  and evoked release events originate from the same vesicular pathway in presynaptic neurons\cite{Bauerfeind1994}. The relation of evoked and spontaneous neurotransmitter releases and how they are distributed spatially have not been precisely studied due to the lack of direct experimental measurement.  In 2008, David Zenisek found some evidence for the spatial segregation of spontaneous and evoked neurotransmissions that evoked release occurs from ribbon and spontaneous release happens from extraribbon locations in a ribbon-type synaptic terminal, in the goldfish retinal bipolar\cite{Zenisek2008}. Kavalali and colleagues examined the evidence that spontaneous and evoked vesicles originate from different pool of glutamate stores and after releasing, neurotransmitters activate non-overlapping postsynaptic NMDA receptors populations\cite{Sara2005, Atasoy2008, Kavalali2011}. Melom and colleagues showed that even though release probability is not correlated between evoked and spontaneous release of fusion. Neuronal dynamics have two spatially segregated and regulated information channels to induce evoked or spontaneous fusion signals independently\cite{Melom2013}. Peled and colleagues suggested that although individual synapses can participate in both evoked and spontaneous neurotransmitter release, there is a highly well activated synapse with a preference for only one mode of transmission\cite{Peled2014}.
Schneggenburger and colleagues found that separate functions for $Ca^{2+}$ evoked release  and spontaneous transmissions are not necessarily from different origins of two vesicular fusion\cite{Schneggenburger2015}.

From those studies, there have been assumed that spontaneous and evoked process are segregated and regulated independently. But it is unclear how this separation of synaptic currents in NMDA receptors is distributed across individual synapses because of the limited resolution of optical microscopic recording. It is still challenging to show how postsynaptic neurons distinguish evoked and spontaneous neurotransmission and differentially activate postsynaptic signaling. Reese and Kavalali showed that two signals from spontaneous and evoked release are not correlated significantly although spontaneous and evoked release driven NMDA receptor mediated $Ca^{2+}$ transients often occurs at the same synapse\cite{Reese2016}. 

To address this question, we examined the possibility that spontaneous and evoked neurotransmission using the mathematical model improved from our previous study\cite{Atasoy2008}. We aimed to address the biophysical constraints that may determine independent activation on N-methyl-D-asparate (NMDA) receptor mediated currents in response to evoked and spontaneous glutamate molecules releases. 

Identifying the mechanism of activation of NMDA receptors is important since the NMDA receptors (NMDARs), a family of L-glutamate receptors, are a main target for cognitive enhancement and plays a significant role in neural plasticity including long-term potentiation and long-term depression\cite{Newcomer2000,Collingridge2013,Kumar2015}. These explain the importance of NMDARs for learning and memory. Recent studies also indicated an age-associated NMDA receptor hypofunction and memory impairment are linked each other, and it provides evidence age-associated may enhance oxidative stress\cite{Kumar2015}. 

In this way, we described the boundaries of different factors, including size and geometry of synaptic cleft, the neurotransmitter release rate s from presynaptic terminals, different rates glutamate mobility that permit independent synaptic events in the limited space of a synapse cleft. The geometry of synapse, especially for a small size, has not been described clearly with the limitation of microscopy. In order to simulate two modes of neurotransmission, we assumed that spontaneous and evoked release with different factors including a size, geometry of synaptic cleft, different glutamate mobility rate in the cleft, and the release rate of neurotransmitters from presynaptic terminals.

%%%%%%%%%%%%%%%%%%%%%%%%%%%%%%%%%%%%%%%%%%%%%%%%%%%%%%%%%%%%%%%%%%%%%%%%%%%%%%%%%%
%%%%%%%%%%%%%%%%      Mathematical Modeling              %%%%%%%%%%%%%%%%%%%%
%%%%%%%%%%%%%%%%%%%%%%%%%%%%%%%%%%%%%%%%%%%%%%%%%%%%%%%%%%%%%%%%%%%%%%%%%%%%%%%%%%
\section{Methodology}
%\subsection{Geometric domain and model assumptions }\label{sec2}

\subsection{Physical setting}
The computational model is calculated by the diffusion and kinetics in a cubic domain $\Omega$ of $1000 \mbox{nm} \times 1000 \mbox{nm} \times 1000 \mbox{nm} $ excluding synaptic terminals. Let S matrix for boundary conditions,

 $$ S = S(x,y,z)=
\begin{cases} 
  1,  & \mbox{ within $\Omega$ }\\
  0, & \mbox{otherwise }
\end{cases}.$$

\begin{figure}[h]%
\centering
\includegraphics[width=0.9\textwidth]{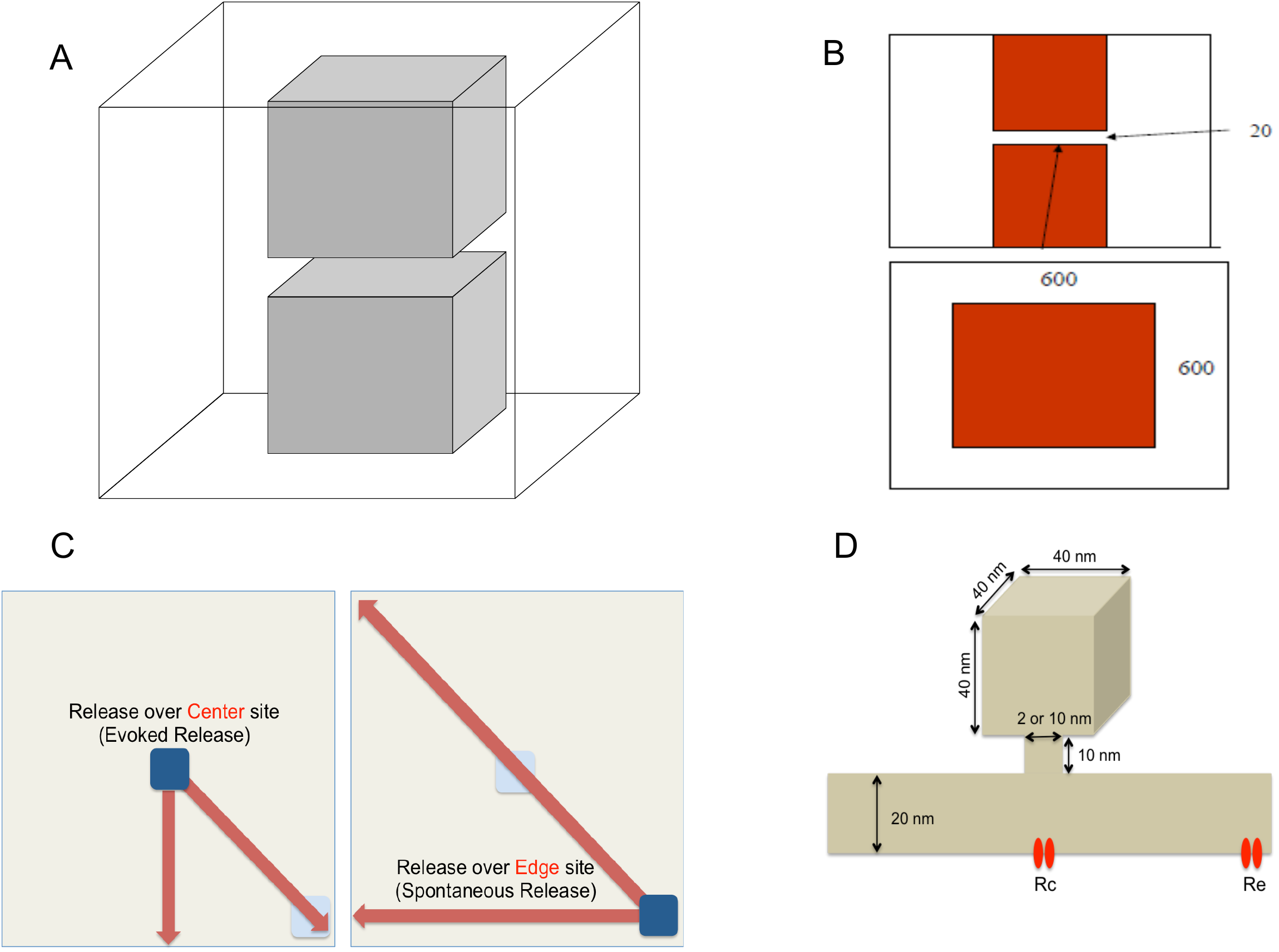}
\caption{\textbf{Geometric domain for computational model in synaptic glutamate diffusion: } \textbf{A.} Illustrates the computational domain in 3-D in the transparent portion, inside a large cube. \textbf{B.}  Top: a side-view of \textbf{A}  shows a cleft of 20 nm height separating two terminals; bottom: a top-view of \textbf{A} shows the cleft surface (in color, dimension $600 \mbox{nm} \times 600 \mbox{nm}$) of a large synapse. \textbf{C.} Two directions that are tracked peak open probabilities at each location on postsynaptic when glutamate molecules release from the center($R_C$, maybe evoked release?) and release from the edge($R_E$, may be spontaneous release?). \textbf{D.} Geometric vesicle fusion pore for computational model in synaptic glutamate diffusion.}\label{fig1}
\end{figure}

 The presynaptic and postsynaptic terminals (two $600 \mbox{nm} \times 600 \mbox{nm}$  surface areas facing each other) inside $\Omega$ are non-permeable by glutamate molecules and are excluded from computation. A cleft of 20 nm heights separates the presynaptic and postsynaptic terminals in the base model. We assumed that size of synapse is  $600 \mbox{nm} \times 600 \mbox{nm} \times 20 \mbox{nm}$ as for the base model. Geometry for a synapse is a three dimensional array, as shown Figure~\ref{fig1}A,~\ref{fig1}B. 
 
%\item Cleft dimensions : 600nm x 600nm x 20nm

For the base model (600nm by 600nm cross-section), we assumed the NMDA receptor density of 40 $\mu m^2$. On the postsynaptic cleft, there are evenly distributed 25 spots, arranged in 5 by 5 array using row-major ordering. The concentration time course and $P_{open}$ are calculated at each receptor. With our assumption, a release events near the center represents evoked transmission, whereas a fusion event at the periphery of the postsynaptic corresponds to spontaneous receptor\cite{Atasoy2008}. Two representative locations of receptors are used as release sites, $R_C$ for central release(released from the center site) and $R_E$ for release at the edge(released from the edge site) as shown in Figure ~\ref{fig1}C. 

We illustrated the situation where evoked glutamate release at the center activates evoked receptor kinetics at center and spontaneous receptor kinetics at edge in Figure~\ref{fig2}A; analogously, spontaneous glutamate release at the cleft edge activates both evoked receptor kinetics at center and spontaneous receptor kinetics at edge in Figure~\ref{fig2}B. For a large size synapse under normal conditions (assumed in the base model), the two signals are independent as central versus edge release yield to synaptic currents that differs in the amplitude (or open probability) of 10 magnitudes. We begin with the base model with instantaneous release of 4000 glutamate molecules.

Another assumptions were considered the limited rate release with additional components of a vesicle and a fusion pore inside the presynaptic terminal as shown in Figure~\ref{fig1}D. The vesicle size is $40 \mbox{nm} \times 40 \mbox{nm}$, the fusion pore size mimics release speed. We tested the release of glutamate molecules from a 10 nm diameter fusion pore for a typical release and one narrow 2 nm diameter fusion pore as a slower release.

\subsection{Mathematical Modeling}
\subsubsection{Diffusion model from presynaptic neuron to synaptic cleft}\label{sec3}
The classical heat diffusion equation was used in the glutamate release process on the presynaptic terminal. The heat diffusion model is reasonable because of the size of glutamate molecules, and relatively large numbers of the molecules being release as one time. We assumed that 4000 glutamate molecules release out from a point source\cite{Nielsen2004, Atasoy2008}. With an estimated current dose-response profiles obtained from measurements\cite{Anson1998, Patneau1990}, it was widely believed that the glutamate-binding sites become saturated after each synaptic vesicle released\cite{Frerking1995}, and the estimated high glutamate concentration in the cleft after the release of a single synaptic vesicle is about 1-5mM\cite{Clements1996, Diamond1997}.  This is consistent with simulated values using Equation~\ref{eqn:heat equation} with initial concentration of 4,000 molecules. The governing equation is

\begin{equation}
\frac{\partial{C}}{\partial{t}}=D_{glut}\bigg(\frac{\partial^2{C}}{\partial^2{x}}+\frac{\partial^2{C}}{\partial^2{y}}+\frac{\partial^2{C}}{\partial^2{z}}\bigg),
\label{eqn:heat equation}
\end{equation}
where $[G]=C(x,y,z,t)$ is a glutamate concentration as a function of time and location in the vesicular space, the synaptic cleft, and the external space, whereas $t\in[0,\infty)$ and $(x,y,z)$ is in a open region $\Omega$. The coefficient $D_{glut}$ is the thermal diffusivity.
%$D_{glut} = \frac{k}{C_p \rho }$ is the thermal diffusivity.

\subsubsection{Factors in synapse}

The diffusion coefficients $D_{glut}$  can represent the inhomogeneity of media within the cleft with varying $D_{glut}$. The larger diffusion coefficient value implies for glutamate to flush out to external space quickly. For a typical synapse, the value of $0.4 \mu m^2/ms$ is sufficient to represent typical case of glutamate mobility\cite{Nielsen2004}. However, this may or may not be the case with small synapses, where the evoked and spontaneous releases occur in much closer space. The exact value of  $D_{glut}$ is unknown, we used values ranging from $0.1$ to $0.75 \mu m^2/ms$ as feasible permeability values for synapse. Further we can create multiple zones where $D_{glut}$ could have various values in several zones. In the models, diffusion coefficients $D_{glut}$ were taken depending on the location within or outside the synapses. In base model, we have a value of $D_{glut}=0.4\mu m^2/ms$ within the cleft and $D_{glut}=0.75\mu m^2/ms$ in the external space. 

We considered other constraints to find favorable conditions where the signals can be independent. One way is to assume the components in the cleft to be different compositions. We make the center area to be less diffusive for glutamate molecules than the peripheral region inside cleft. We created an interior zone($\Omega^+$) and an exterior zone($\Omega^-$) as shown in Figure \ref{fig6}A, \ref{fig6}B. $D_{glut}$ could vary zone to zone,  and it represents the inhomogeneity of material components in the cleft. Numerically, explicit finite difference scheme with piecewise continuous coefficients has accuracy of second order $O(h^2)$ for the synaptic diffusion problems\cite{Seo2017, Seo2019}.

Another possibility is to reduce the amount of glutamate molecules released per vesicle in the partial release (called kiss-and-go) scenarios, and/or the glutamate is released in slower release rate due to partial opening of vesicles in the membrane. When simulating the vesicular diffusion, $D_{glut}=0.15\mu m^2/ms$ and $D_{glut}=0.0375\mu m^2/ms$ are taken for 10nm and 2nm fusion pore respectively. 

\subsubsection{Post-synaptic model}
%\subsection{Kinetic model from synaptic cleft to post synaptic neuron}\label{sec4}
Under the glutamate mediation, receptors such as NMDA, AMPA or GABA activate the channels for allowing ion currents. To determine the opening probability, $P_{open}$, of an individual receptor, a state model is proposed based on the maximum likelihood method using experimental data. The current model consisting of three closed and two open states (3C2O) is used by Popescu data\cite{Popescu2004, Atasoy2008}. Our calculation will be based on this 3C2O model. We note that the glutamate concentrations at the receptor locations are included in the reaction rates of two coupled states, $C^M$and $C^U$.

The experimental data indicated the populations of synapses are clustered around 3 groups of distinct characters. We primarily tested our hypotheses on the medium group called M-mode is modeled by   
\begin{equation}
C^{U} \underset{58}{\overset{39[G]}{\rightleftharpoons}}C^{M}\underset{116}{\overset{19[G]}{\rightleftharpoons}}C_1\underset{173}{\overset{150}{\rightleftharpoons}}C_2\underset{2,412}{\overset{902}{\rightleftharpoons}}C_3\underset{1,283}{\overset{4,467}{\rightleftharpoons}}O_{1}\underset{526}{\overset{4,630}{\rightleftharpoons}}O_2,
\end{equation}
The ode systems were built using kinetic model for open probability at each receptor when glutamate molecules releases occur. The open probability is obtained by $P_{open} = O_1(t)+O_2(t) $ \cite{Atasoy2008}. The populations satisfy the following system of ordinary differential equations:
\begin{eqnarray*}
\frac{dC^U}{dt}&=&58C^M-39C(x,y,z,t)C^U,\\
\frac{dC^M}{dt}&=&(116C_1-19C(x,y,z,t)C^M)-(58C^M-39C(x,y,z,t)C^U),\\
\frac{dC_1}{dt}&=&-(116C_1-19C(x,y,z,t)C^M)+(173C_2-150C_1),\\
\frac{dC_2}{dt}&=&(2412C_3-902C_2)-(173C_2-150C_1),\\
\frac{dC_3}{dt}&=&-(2412C_3-902C_2)+(1283O_1-4467C_3),\\
\frac{dO_1}{dt}&=&(526O_2-4630O_1)-(1283O_1-4467C_3),\\
\frac{dO_2}{dt}&=&-(526O_2-4630O_1).
\end{eqnarray*}

\subsubsection{Computational model}

We simulated the process of glutamate release from presynaptic sites into the synaptic cleft by solving the heat diffusion equation and the synaptic transmitter/receptor kinetic process by solving a system of ordinary differential equations. We used the problem-solving environment MATLAB, which provides tools for solving linear equations in a numerical way. MATLAB is efficient to display results visually through graphs in its post-processing\cite{Shampine1997}. In order to solve the heat equation numerically to achieve glutamate concentration, we use on explicit difference method (forward time, centered space) that is implemented in MATLAB codes. 
\begin{eqnarray*}
C^{n+1}_{i,j,k}&=&C^{n}_{i,j,k}+\alpha[S_{i+1,j,k}C^{n}_{i+1,j,k}+S_{i-1j,k}C^{n}_{i-1j,k}+S_{i,j+1,k}C^{n}_{i,j+1,k}\\
&+&S_{i,j-1,k}C^{n}_{i,j-1,k}+S_{i,j,k-1}C^{n}_{i,k,j,k-1}-(S_{i,j,k}C^{n}_{i,j,k}+S_{i-1,j,k}C^{n}_{i,j,k}\\
&+&S_{i,j+1,k}C^{n}_{i,j,k}+S_{i,j-1,k}C^{n}_{i,j,k}+S_{i,j,k+1}C^{n}_{i,j,k}+S_{i,j,k-1}C^{n}_{i,j,k})],
\end{eqnarray*}
where $C^n_{i,j,k} = C(x_i, y_j, z_k, t_n)$ and $\alpha = D_{glut}\frac{\Delta t}{(\Delta x)^2}$. 
For our simulation, we satisfy a condition of $\alpha < (1/2)^3$  known as CFL condition to ensure the scheme is numerically stable\cite{Lewy1928}. We take the space step $\Delta x= 0.01\mu m$ and the time step $\Delta t = C\cdot (\Delta x)^2$, where $C$ is a constant.

The glutamate concentration $C(x,y,z,t)$  simulating was averaged over $10\mu s$ intervals (10 time steps) to reduce computational cost. MATLAB solver ode23s is effective at crude errors and relatively faster than other solvers\cite{Shampine1997}.

\subsection{Measurement for independence}
We denoted the relative ratio function as the ratio of the maximum open probability of distal receptor over the maximum opening probability of the receptor directly opposing the release site (center or edge). Respectively, we call a receptor as  $R_C$, which is directly opposing the evoked release site, and also represents a receptor located the center of postsynaptic terminal. A receptor $R_E$,  directly opposing the spontaneous release site, is  located around the edge of the postsynaptic terminal. The measure(\ref{eq1:measurement}) we define is showing the points where ratio functions are taken for comparison:

\begin{equation}
Measure := \frac{\mbox{Max}(P_{open}) \  \mbox{at distal receptor}}{\mbox{Max}(P_{open}) \ \mbox{at release site}\  (R_C\ \mbox{or}\  R_E)}.
\label{eq1:measurement}
\end{equation}

 When we simulate our model, we fix the denominator as one receptor either $R_C$ and $R_E$. Then we measure Max($P_{open}$) at a receptor along the two directions as shown in Figure~\ref{fig1}C. The Max($P_{open}$) in other locations will be valued between the values obtained in these two directions. If the steepness of function for ratio is relatively large, it implies that the evoked and spontaneous currents have a less chance of having crosstalk. We used the measure (\ref{eq1:measurement}) to estimate the independency with considering effective conditions for independency such as a synaptic size, diffusion inhomogeneity, and  fusion pore. When glutamate vesicle releases over the center receptor in Figure~\ref{fig1}C(left), for the receptors of equal distance to release, the diagonal direction shows the largest $P_{open}$ and vertical (or horizontal) direction shows the smallest given the same distance. In this process the Piecewise Hermite Cubic Interpolation (PHCI) was used to approximate the probable values to preserve monotone or convex curve of the function\cite{Hyman1983}.

%%%%%%%%%%%%%%%%%%%%%%%%%%%%%%%%%%%%%%%%%%%%%%%%%%%%%%%%%%%%%%%%%%%%%%%%%%%%%%%%%%
%%%%%%%%%%%%%%%%        Results              %%%%%%%%%%%%%%%%%%%%
%%%%%%%%%%%%%%%%%%%%%%%%%%%%%%%%%%%%%%%%%%%%%%%%%%%%%%%%%%%%%%%%%%%%%%%%%%%%%%%%%%
\section{Results}\label{sec5}
\begin{figure}[h]%
\centering
\includegraphics[width=0.9\textwidth]{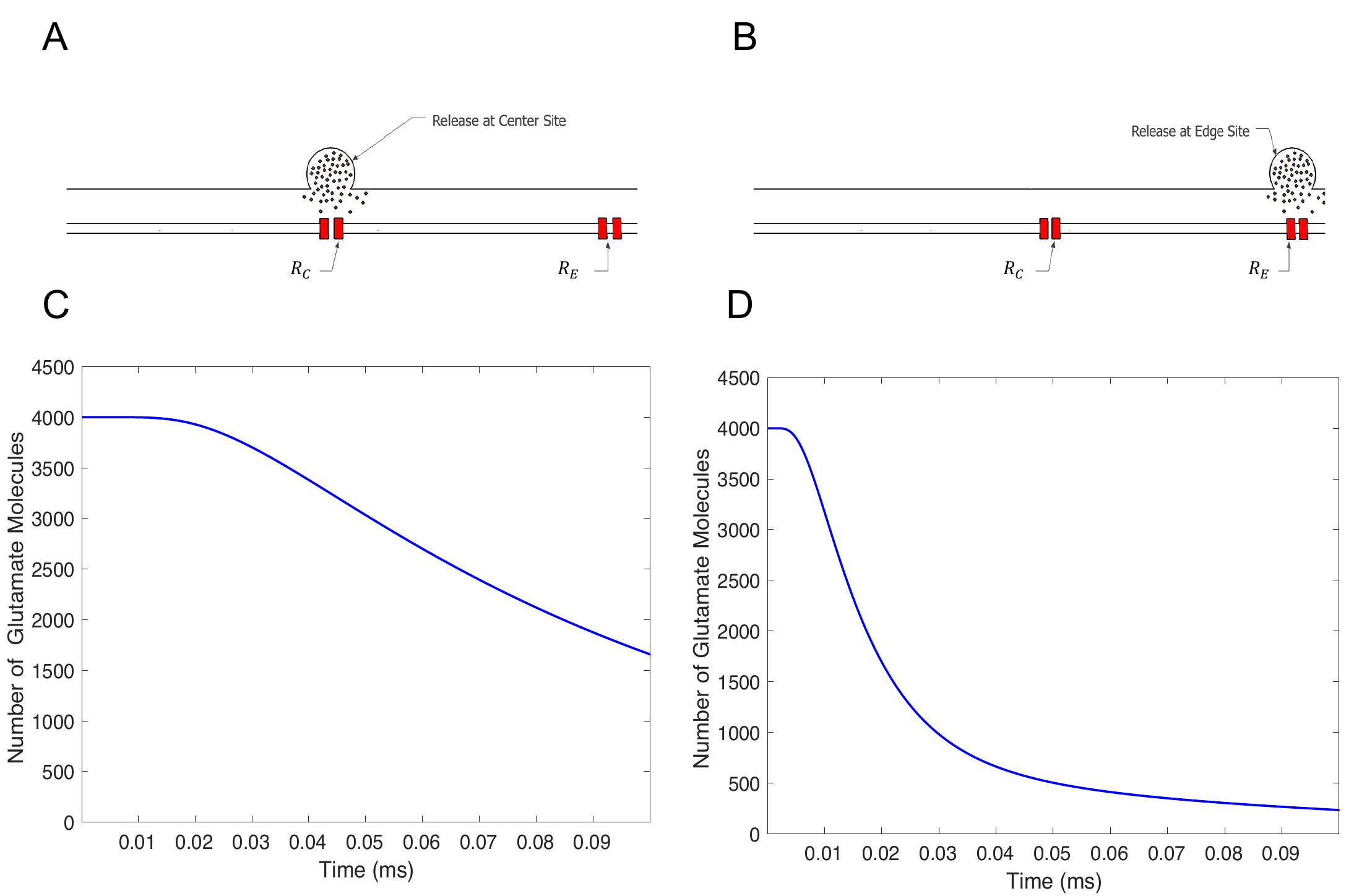}
\caption{\textbf{A.} Hypothesis that glutamate molecules release over the center at the cleft (a typical location is directly opposing to the center on the postsynaptic neuron, called $R_C$). \textbf{B.} Hypothesis that glutamate molecules release over the center at the cleft (a typical location is directly opposing to the center on the postsynaptic neuron, called $R_E$). \textbf{C.} The total number of molecules inside of a cleft is depicted as a function of time, in an instantaneous release of 4000 glutamate molecules at $R_C$ for glutamate molecule release at the center. \textbf{D.} The total number of molecules inside of a cleft is depicted as a function of time, in an instantaneous release of 4000 glutamate molecules at $R_E$ for glutamate molecule release at the edge.}\label{fig2}
\end{figure}

We assumed the situation where evoked glutamate release activates evoked receptor kinetics at center and spontaneous receptor kinetics at edge as shown in Figure~\ref{fig2}A,~\ref{fig2}B. Spontaneous glutamate release at the cleft edge activates evoked receptor kinetics at center and spontaneous receptor kinetics at edge.

First, we simulated the base model with an instantaneous release of 4000 glutamate molecules. Figure~\ref{fig2} shows the total number of molecules within the cleft decreases according to the time when center release and edge release occur, respectively. It takes about 0.08ms to clear a half of the molecules out of the cleft and the population decays in exponential trends with decay constants $8.808 \times 10^{-5}$ and $4.526 \times 10^{-4}$ respectively, as shown Figure~\ref{fig2}C,~\ref{fig2}D . It implies that an edge release makes glutamate molecules flushed out much quicker than that of center release.

%Figure \ref{fig3} shows concentrations of glutamate at each receptor location when 4000 glutamate molecules are released instantaneously at $R_C$ near the center of the presynaptic terminal and released at $R_E$ near the edge of the presynaptic terminal respectively (See Figure \ref{fig3}A,~\ref{fig3}B). Clearly, we can verify that the location over $R_C$ has the highest glutamate concentration among 25 spots when center release occurs. Analogously, the peak glutamate concentration appears at $R_E$  receptor when an edge release occurs. 

%\begin{figure}[h]%
%\centering
%\includegraphics[width=1 \textwidth]{figure3.pdf}
%\caption{ Glutamate concentrations time series for 0.1 $\mu s$, plotted at 16 receptor locations when 4000 glutamate molecules are released at R6 near the center of the presynaptic terminal(\textbf{A}), and when 4000 glutamate molecules are released instantaneously at R16 near the edge of the presynaptic terminal(\textbf{B}), respectively.}\label{fig3}
%\end{figure}

\subsection{Effect of size of synapses }

\begin{figure}[h]%
\centering
\includegraphics[width=1 \textwidth]{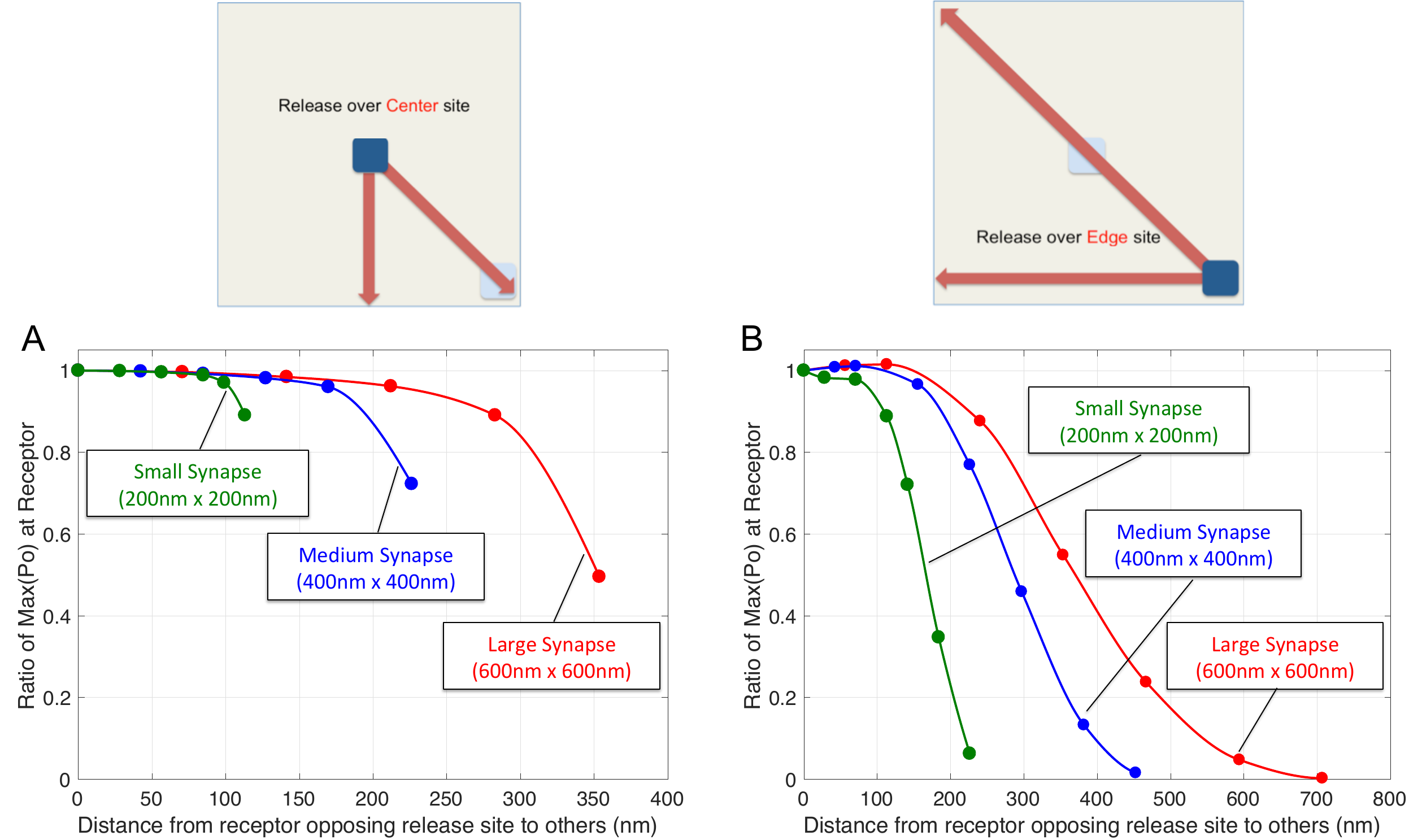}
\caption{Ratios of maximum NMDA receptor opening probabilities($P_{open}$) as functions of receptor distance for three sizes of synapse as calculated by the kinetics equation, when glutamates are released above the center location(\textbf{A}) and the edge location(\textbf{B}), respectively. }\label{fig3}
\end{figure}

We used the model to calculate and compare three sizes of synapses as large synapse, medium synapse, and small synapse. Then we set areas of synapse as $600\mbox{nm} \times 600\mbox{nm}$ for large synapse, $400\mbox{nm} \times 400\mbox{nm}$ for medium size, and $200\mbox{nm} \times 200\mbox{nm}$ for small synapse respectively. We estimated the minimum distance between two sets of NMDA receptors to have possible less crosstalk. The ratios of maximum opening probabilities are plotted in Figure \ref{fig3}A and \ref{fig3}B. 

 Based on the criteria from our previous research \cite{Atasoy2008, Reese2016}, we assumed 5-fold(0.5, see in Figure \ref{fig3}) is the threshold ratio as a good indicator for independent currents. However, it is hard to assure independency of evoked and spontaneous neurotransmission for medium ($0.16\mu m^2$) and small synapses ($0.04\mu m^2$) in our current base model. For medium synapse ($400\mbox{nm} \times 400\mbox{nm}$), if two forms of release locate towards opposite corners of the synaptic cleft, we might achieve sufficiently low level of crosstalk and possibly obtain the independence of spontaneous and evoked neurotransmissions \cite{Atasoy2008}. For small synapse $200\mbox{nm} \times 200\mbox{nm}$, there are apparently have more crosstalk between evoked and spontaneous releases.

\subsection{Effect of diffusion inhomogeneity in synaptic cleft}

\begin{figure}[h]%
\centering
\includegraphics[width=1 \textwidth]{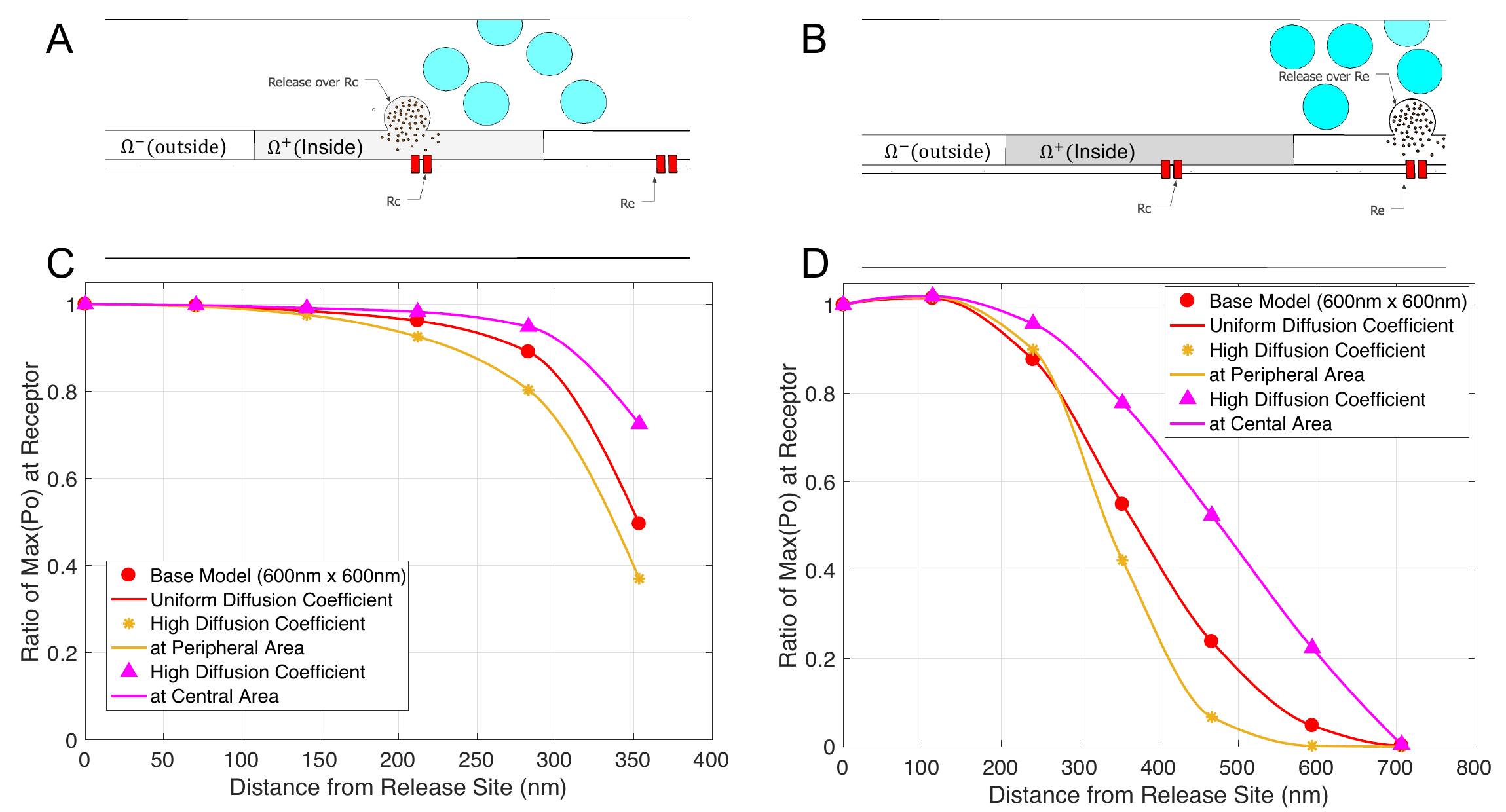}
\caption{\textbf{A}, \textbf{B}. The synaptic cleft space is divided into two zones. The diffusion coefficients assume two values in two regions($\Omega^+$ and $\Omega^-$) respectively, which represent slow and fast motion of neurotransmitters in different material composition of the cleft space. Front view of diffusion process with two different zones when center release(A) or edge release(B) occur. \textbf{C}, \textbf{D}. Ratio of Max($P_{open}$) as a function of receptor distance for diffusion inhomogeneity(base, high affinity center, and high affinity edge) when center release(C) or edge release(D) occur.}\label{fig4}
\end{figure}

We considered an assumption that the background medium for diffusion of glutamate molecules may be different depending on the location inside the cleft. We created an interior zone($\Omega^+$) and an exterior zone($\Omega^-$) as shown in Figure \ref{fig4}A, \ref{fig4}B. $D_{glut}$ could vary zone to zone,  and it represents the inhomogeneity of material components in the cleft. 

In Figure \ref{fig4}, among three lines for both center(evoked) and edge(spontaneous) release, the high affinity center model has the sharper downward slope when the distance is apart. However, from 0 to 150nm range of distance, all three lines are virtually constants as shown in Figure \ref{fig4}, and this implies that this high affinity center is not enough for a small synapse to house the independent currents from two modes transmissions. 

\subsection{Effect of narrow fusion pores}

\begin{figure}[h]%
\centering
\includegraphics[width=1 \textwidth]{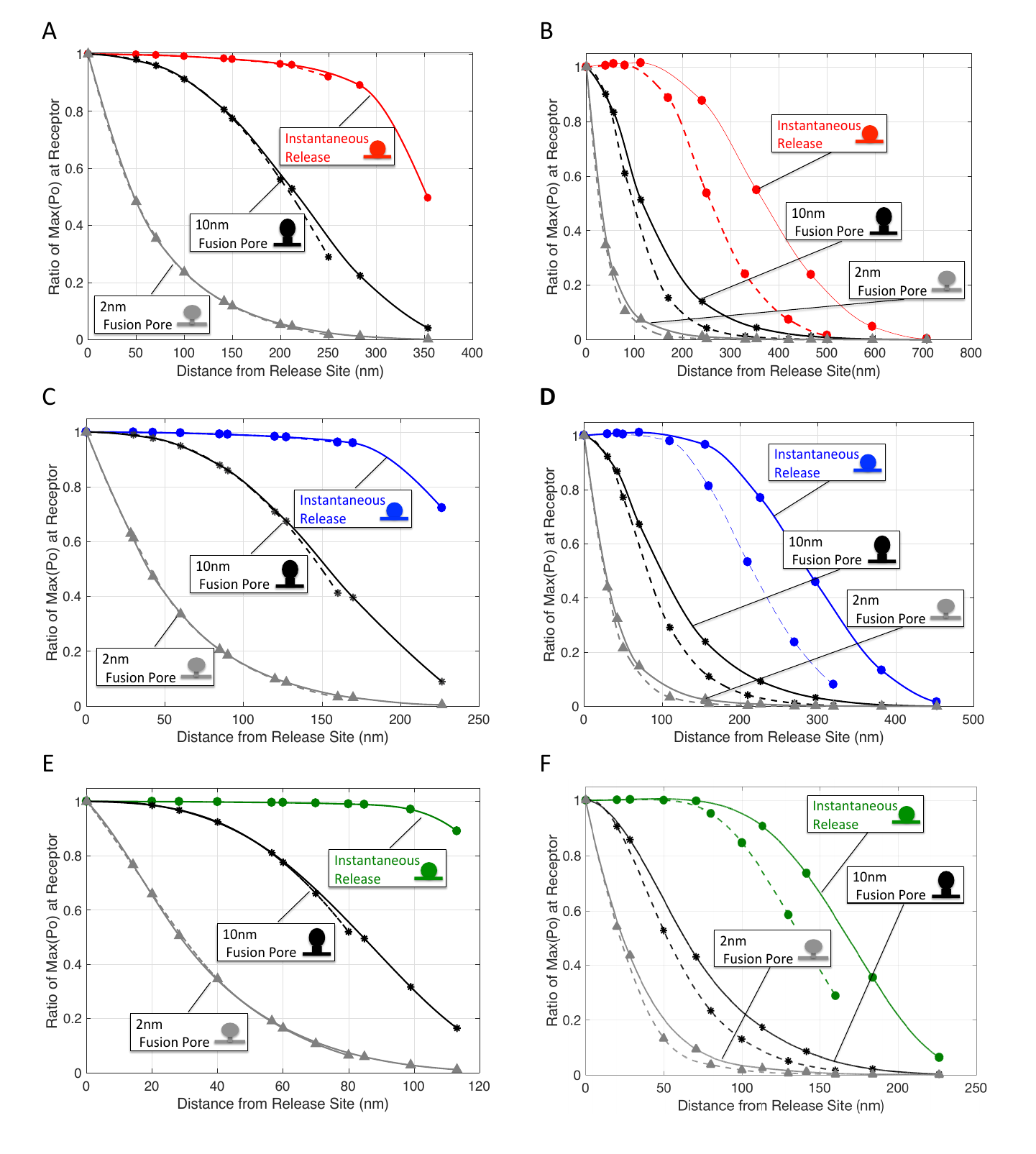}
\caption{ Ratios of maximum NMDA receptor opening probabilities as functions of receptor distance for different release speed (slow, 2 nm fusion pore - $\triangle$, regular, 10 nm fusion pore- $\ast$, and instantaneous -$\circ$) of glutamate vesicle release. The open probabilities were calculated by the kinetics equation, when glutamates are released above the center location in \textbf{A}, \textbf{C}, \textbf{E} and the edge location in (\textbf{B}, \textbf{E}, \textbf{F}) respectively.  The sizes of the synapses are $600 \mbox{nm} \times 600 \mbox{nm}$ in (\textbf{A}, \textbf{B}), $400 \mbox{nm} \times 400 \mbox{nm}$ in (\textbf{C}, \textbf{D}), and $200 \mbox{nm} \times 200 \mbox{nm}$ in (\textbf{E}, \textbf{F}), respectively. }\label{fig5}
\end{figure}

One possibility is to reduce the amount of glutamate molecules released per vesicle in the partial release (called kiss-and-go) scenarios, and/or the glutamate is released in slower release rate due to partial opening of vesicles in the membrane. When simulating the vesicular diffusion, $D_{glut}=0.15\mu m^2/ms$ and $D_{glut}=0.0375\mu m^2/ms$ are taken for 10nm and 2nm fusion pore respectively. As shown in Figure \ref{fig5}, the limited vesicle fusion rate has impact on reducing the crosstalk of two currents from center and edge releases than that of instantaneous model (our base model). They did not impact as much to perturb the independent signaling of synapse on large synapses, although slower releases did promote more independence. Figure \ref{fig5}\textbf{E} shows that for small synapse (200nm x 200nm) with instantaneous release, the ratio of maximal opening probability is close to 1 and is consistent up to 90nm when center release occurs, this implies that they have high probability of having crosstalk of two neurotransmissions. In fact, for small size of synapse, there is not much difference between the maximal opening probability at $R_C$ and $R_E$ when center release or edge release occurs. This limited vesicle fusion model benefits small size synapses substantially in achieving independent signaling. This can be verified in the graph of Figure \ref{fig5}\textbf{E} that as glutamate molecules release through 10nm or 2nm vesicle fusion pore, the ratio of maximal opening probability has the sharper downward slope when distance goes away. The ratio achieves 10-fold reduction at 90nm distance, giving plausibility for independent currents from two transmissions.  For edge release, limited vesicle fusion is a redundant factor for activating two independent fusions since the ratio drastically decreases and becomes close to zero at 100nm distance for all size of synapses as shown in Figure \ref{fig5}\textbf{B}, \ref{fig5}\textbf{D}, and \ref{fig5}\textbf{F}. This is because when two release sites are located to the two opposite corners, then most of glutamate molecules flush out of the cleft very fast and rarely reach to the other site of receptor. 

\subsection{Simulation results compared with experimantal finding of PSD-95 enrichment }
An alignment of presynaptic and postsynaptic nanoscale subdomains, called nanocolumn was proposed. Evoked fusion occurs in confined areas by protein gradient with higher local density of Rab3-interacting molecules (RIM) within the presynaptic active zone\cite{Tang2016}. These RIM nanoclusters align with concentrated postsynaptic receptors. Evoked neurotransmitter release prefer to occur at sites directly opposing postsynaptic receptor guided by the nanoachitecture of the active zones. They estimated that majority(72-82\%) of evoked signals arose from single vesicle fusion. The concentration of vesicle priming proteins in nanoclusters prefers to evoked fusion in the subregion of the active zone. Three RIM 1/2 nanoclustes and three PSD-95(Post Synaptic Density) nanoclusters are well aligned for two pairs and not aligned for one pair. They used two independent approaches to estimate the relationship between active zone and postsynaptic density(PSD) protein distributions. In order to figure out of the trend, they measured RIM 1/2 localization densities as a function of radial distance from the centres of PSD-95 nanoclusters as translated across the synaptic cleft. Similarly, they estimated PSD-95 protein enrichment densities as a function of the center of RIM 1/2 nanoclusters. 

 \begin{figure}[h]%
\centering
\includegraphics[width=1 \textwidth]{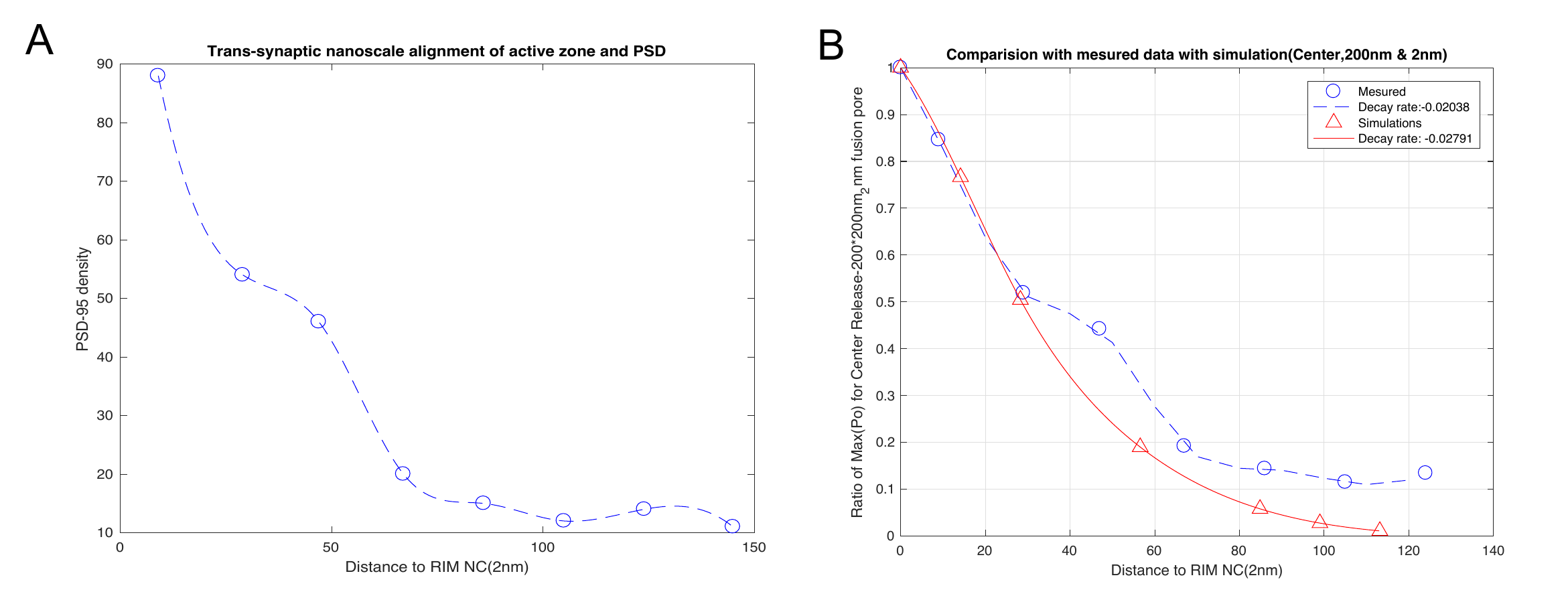}
\caption{\textbf{A.} PSD-95 enrichment as a function of distance relative to RIM 1/2 nanoclusters \cite{Tang2016}. \textbf{B.} Comparsion with the rescaled data with PSD-95 Enrichment from \textbf{A}(red) and simulation results (center, 200nm \& 2nm) with the measurement (blue, dashed). }\label{fig6}
\end{figure}

Through this results, they defined an enrichment index as the average molecular density of the opposed protein (n=265-272) within a 60nm radius from the nanocluster center. They verified from this results if synapses are trans-synaptically aligned on the nanoscale level, the distribution of protein on side of the synapse may predict protein density in the opposing neuron. The experimental findings seemed to be relevant to our mathematical results, so we compared the measured data with the results from our mathematical measurement for small synapse(200nm by 200nm). The data \cite{Tang2016} was obtained and fitted with the exponential model using MATLAB. Figure \ref{fig6}A has trends of exponentially decreasing. The measured data of PSD-95 enrichment on the postsynaptic neuron as a function of distance relative to the center of  RIM 1/2 nanocluster on presynaptic terminal  is approximating to the exponential function as below: 
\begin{equation}
f_{PSD-95}(x) = a_1 \cdot \exp (b_1\cdot x)
\label{eq1}
\end{equation}
The coefficients with 95 \% confidence bounds of $a_1$ and $b_1$ were estimated to 104 (87.06, 121) and -0.02038 (-0.02518, -0.01558), respectively.  

We tested our simulation and suggested that we may obtain the independency of two currents from evoked and spontaneous neurotransmitter release when if small synapses have a geometry of a narrow fusion pore (2nm)(Figure \ref{fig5}E, \ref{fig5}F). The ratio achieves 10-fold reduction at 90nm distance, giving plausibility for independent currents from two modes of transmissions. This corresponds to the result of an enrichment index within a 60nm radius from the nanocluster centre. Thus, we also consider the graph of ratio function of peak opening probability as a function of distance from the center synapse for small synapse and evoked narrow fusion pore release in Figure \ref{fig5}E. Then we fit the graph to the exponential model similarly. The coefficients with 95 \% confidence bounds of $a_2$ is 1.044 (0.9322, 1.156) and $b_2$ is approximating to -0.02791 (-0.03409, -0.02171).  We obtained an exponential function for our simulation,
\begin{equation}
f_{2nm-sim}(x) = a_2\cdot \exp (b_2 \cdot x).
\label{eq2}
\end{equation}

The coefficients of $f_{PSD-95} ($\ref{eq1}) were rescaled and plotted the rescaled function of $f_{PSD-95} ($\ref{eq1}) with the function of $f_{2nm-sim}$(\ref{eq2}) together as shown in Figure \ref{fig6}. 

%%%%%%%%%%%%%%%%%%%%%%%%%%%%%%%%%%%%%%%%%%%%%%%%%%%%%%%%%%%%%%%%%%%%%%%%%%%%%%%%%
%%%%%%%%%%%%%%%%        Discussion              %%%%%%%%%%%%%%%%%%%%
%%%%%%%%%%%%%%%%%%%%%%%%%%%%%%%%%%%%%%%%%%%%%%%%%%%%%%%%%%%%%%%%%%%%%%%%%%%%%%%%%%

\section{Conclusion}

In this study, we aimed to address the biophysical constraints that may determine independent activation on N-methyl-D-asparate (NMDA) receptor mediated currents in response to evoked and spontaneous glutamate molecules releases though simulating a mathematical model. In this way, we described the boundaries of different factors, including size and geometry of synaptic cleft, the neurotransmitter release rate s from presynaptic terminals, different rates glutamate mobility that permit independent synaptic events in the limited space of a synapse cleft. The previous research has shown the evidence that spontaneous and evoked vesicles originate from different pool of glutamate molecules stores, and after releasing, neurotransmitters activate non-overlapping postsynaptic NMDA receptors populations for a large synapse\cite{Sara2005, Atasoy2008, Kavalali2011}. However, it is still unclear the mechanism of evoked and spontaneous neurotransmission that activate on postsynaptic terminals.Thus, we constructed a mathematical model of two modes of neurotransmission, and simulated spontaneous and evoked release process with different factors including a size, geometry of synaptic cleft, different glutamate mobility rate in the cleft, and the release rate of neurotransmitters from presynaptic terminals. 

\subsection{Discussion}
The geometry of synapse, especially for a small size, has not been described clearly with the limitation of microscopy. For large synapses, experimental evidence supports the hypothesis that spontaneous and evoked currents are resulted from glutamate vesicle releases in different pools and after releasing, neurotransmitters activate non-overlapping postsynaptic NMDA receptors populations \cite{Reese2016, Atasoy2008}. Based on this hypothesis, the maximal open probability is most sensitive to the distance from the release sites. Thus, the independence of currents from two modes of releases mainly resulted from a large ratio of peak open probability at evoked receptors(maybe occurred at the center) or spontaneous receptor(maybe occurred around the edge) over the different release sites($R_C$ or $R_E$). This is a measurement for independence, when we tested the model on a few conditions such as glutamate release site/receptor distance change or glutamate release speed change, as well as other changes in geometry, diffusion inhomogeneity. Based on the criteria from our previous research \cite{Atasoy2008, Reese2016}, we assumed 5-fold(0.5, see in Figure \ref{fig3}) is the threshold ratio as a good indicator for independent currents. However, it is hard to assure independency of evoked and spontaneous neurotransmission for medium ($0.16\mu m^2$) and small synapses ($0.04\mu m^2$) in our current base model. For medium synapse ($400\mbox{nm} \times 400\mbox{nm}$), if two forms of release locate towards opposite corners of the synaptic cleft, we might achieve sufficient low level of crosstalk and possibly obtain the independence of spontaneous and evoked neurotransmissions \cite{Atasoy2008}. For small synapse $200\mbox{nm} \times 200\mbox{nm}$, there are apparently have more crosstalk between evoked and spontaneous releases. 

We propose the two possible scenarios to find favorable conditions where the signals from two neurotransmissions could be independent on the postsynaptic neurons. First,  it is to assume the components in the cleft to be different compositions. We constructed the model with the center area to be less diffusive for glutamate molecules than the peripheral region inside cleft. In Figure \ref{fig4}, among three lines for both center(evoked) and edge(spontaneous) release, the high affinity center model has the sharper downward slope when the distance is apart. This is compatible with the recent experimental results that the evoked release is guided by the protein gradient and prefer to occur in confined area with in high local density of Rab3-interacting molecule(RIM)\cite{Tang2016} in center area. However, from 0 to 150nm range of distance, all three lines are virtually constants as shown in Figure \ref{fig4}A, and this implies that this high affinity center is not enough for a small synapse to house the independent currents from two modes transmissions. The second scenario is to reduce the amount of glutamate molecules released per vesicle in the partial release (called kiss-and-go) scenarios, and/or the glutamate is released in slower release rate due to partial opening of vesicles in the membrane. The instantaneous release of 4000 glutamate molecules is an approximation for the actual situation. The release of glutamate from vesicles in presynaptic terminals is a complex process that includes elevation of Ca2+, binding of SNARE protein to the membrane and a sequence of events of biochemical reactions\cite{Sudhof2011}. As shown in Figure \ref{fig5}, the limited vesicle fusion rate has less impact on the crosstalk of two currents from center and edge releases than that of instantaneous model (our base model). They did not impact as much to perturb the independent signaling of synapse on large synapses, although slower releases did promote more independence. Slow release model with narrow fusion model is fitted with a small size of synapse and substantially more in achieving independent signaling. Figure \ref{fig5}E and \ref{fig5}F show that small synapse (200nm by 200nm) might not have independent signaling when evoked glutamate release occurs instantaneously because evoked and spontaneous release sites are not far away from each other and thus they have high probability of having crosstalk between each other. In fact, there is not much difference between the peak opening probabilities at a receptor for evoked release and the one for spontaneous release. This could be verified in the graph of  Figure \ref{fig5}E, the ratio of two maximum open probabilities is close to 1, and the peak open probability in NMDA receptors is consistent up to 90nm far from the receptor molecules opposing evoked release site. However for small synapses, as glutamate release through 10nm and 2nm vesicle fusion pore, the open probability ratio decreases more drastically and becomes close to zero, and in 2 nm pore, the ratio achieves 10-fold reduction at 90nm distance, giving plausibility for independent currents from two transmissions. Thus, we suggested that we may obtain the independency of two currents from evoked and spontaneous neurotransmitter release when if small synapses have a geometry of a narrow fusion pore (2nm)(Figure \ref{fig5}F). Simulation results showed that the ratio achieves 10-fold reduction at 90nm distance, giving plausibility for independent currents from two modes of transmissions. This corresponds to the result of an enrichment index within a 60nm radius from the nanocluster centre\cite{Tang2016}. We then focused on the graph the a ratio function of peak opening probability as a function of distance from the center synapse for small synapse and evoked narrow fusion pore release in Figure \ref{fig5}E. The coefficients of $f_{PSD-95}$(\ref{eq1}) were rescaled and plotted with the function of $f_{2nm-sim}$(\ref{eq2}) together(Figure \ref{fig6}). The results are well-matched to each other, and this indicates that small synapses might be conducted dynamic functional modules and possibly hold the segregation of sites for spontaneous versus evoked neurotransmission within individual synapses. Tang and colleages found how distribution of presynaptic vesicle sites corresponds to the receptors in the postsynaptic neuron. They supported that action-potential-evoked fusion is guided by scaffolding proteins, called nanocolumn, which were likely aligned near the centre of synapses than near the edge. Also this nanocolumn theory proposed that the active zone for evoked vesicle fusion occurs at sites directly opposing postsynaptic receptors\cite{Tang2016}. These findings indicate that the segregation of sites for spontaneous and evoked neurotransmission with nanoscale subdomains connecting presynaptic and postsynaptic terminals. Therefore, it is possible even for small synapses to possess this neural functional dynamics.

This study had several limitations. First, the diffusion process may not regard inhomogeneous structure in the cleft in the current study, and we may consider the diffusion process in the cleft by adding an advection term in our future study. Secondly, the small number of glutamate molecules in the vesicle challenged in implementing mathematical models of the release process. In addition, the scale of neurotransmission is hundreds nanometer ranges, and it is laid between classical mechanics and quantum mechanics.

In conclusion, we developed a mathematical model to analyze those results for independent signaling of spontaneous and evoked glutamate releases in a single synapse, comparing with the experimental and theoretical prediction. We defined a measurement of independency and set a criterion of a 5 fold ratio as a reasonable boundary for the independence. From those results we suggested two possibilities for small synapses to be less crosstalk from spontaneous and evoked neurotransmitter currents on postsynaptic terminals. It validated through comparisons with experimental findings. The results were well-matched to each other and this implies that small synapses might be conducted of dynamic functional modules and possibly hold the segregation of sites for spontaneous versus evoked neurotransmission within individual synapses. These results show our mathematical modeling may refer to the neuroscience questioning of distribution and separation of NMDA receptors for two independent spontaneous and evoked release process.

\backmatter

\bmhead{Acknowledgments}
We would like to thank Prof. Ege Kavalali for introducing this problem to us and for his advice and discussion in neuroscience. This research was partially supported by the National Research Foundation of Korea(NRF) grant funded by the Korea government, MSIP; Ministry of Science, ICT \& Future Planning (2020R1G1A1A0100746911).


%\bibliography{plain}
%\bibliography{sn-bibliography2022}% common bib file
%\nocite{*}  
%\input reference.tex
%\input sn-article.bbl
%\bibliographystyle{plain}
%\bibliographystyle{unsrt}
%\bibliography{test}
%\bibliography{alpha}
%\bibliographystyle{IEEEtran} % use IEEEtran.bst style                % list all refs in database, cited or not
%\bibliography{refs}           % bib database file refs.bib
%\bibliography{IEEEabrv,IEEEexample1}
%\bibliography{sn-sample-bib.tex}
%\nocite{*}   
   
%%%%%%%%%%%%%%%%%%%%%%%%%%%%%%%%%%%%%%%%%%%%%%%%%%%%%%%%%%%%%%%%%%%%%%%%%%%%%%%%%%
%%%%%%%%%%%%%%%%         Biographical Statement               %%%%%%%%%%%%%%%%%%%%
%%%%%%%%%%%%%%%%%%%%%%%%%%%%%%%%%%%%%%%%%%%%%%%%%%%%%%%%%%%%%%%%%%%%%%%%%%%%%%%%%%
%\thebiography

%\input sn-bib2022.tex%

\end{document}